# The social microbiome: the missing mechanism mediating the sociality-fitness nexus?

Alice Baniel[1,2], Marie J. E. Charpentier[3]

**Affiliations:**

[1] Center for Evolution and Medicine, Arizona State University, Tempe, AZ, 85281, USA.

[2] School of Life Sciences, Arizona State University, Tempe, AZ, 85287, USA.

[3] Institut des Sciences de l'Évolution de Montpellier UMR5554, CNRS, IRD, EPHE, Université de Montpellier, Montpellier, France.

**Correspondence:**

Alice Baniel (alice.baniel@gmail.com)

Marie Charpentier (marie.charpentier@umontpellier.fr)

**ABSTRACT**

In many social mammals, early life social adversity and social integration largely predict individual health, lifespan and reproductive success. Efforts in identifying the physiological mechanisms mediating the relationship between the social environment and individual fitness have so far concentrated on socially-induced stress, mediated by alterations in neuroendocrine signaling and immune function. Here, we propose a much-needed alternative mechanism relying on microbially-mediated effects: social relationships with conspecifics, both in early life and adulthood, might strongly contribute both to the transmission of beneficial microbes and to diversifying host microbiomes. In turn, more valuable and diverse microbiomes would promote pathogen resistance and optimal health and thus translate into positive fitness outcomes. This mechanism relies on two emerging findings from empirical studies, namely that microbiomes (i) are largely socially transmitted via vertical and horizontal routes, and (ii) play a pervasive role in host development, physiology, metabolism, and susceptibility to pathogens. We suggest that the social transmission of microbiomes has the potential to explain the sociality-fitness nexus, to a similar - or even higher - extent than chronic social stress, in ways that have yet to be studied empirically in social mammals.

**TEXT**

The quality of the social environment is one of the strongest predictors of health and longevity in humans[1,2], a conspicuous relationship also observed in many social mammals – with effect sizes of strikingly high amplitudes[3]. Early-life social adversity, social integration, and social status are the three major aspects of the social environment strongly predicting fitness[3]. In all mammalian orders, individuals that experience adverse social conditions in early life, such as maternal loss[4–6], are poorly social integrated throughout their lifetime[7–11], or – to a lesser

extent – acquire low social status[10,12,13], display considerably higher offspring mortality and shorter life spans than individuals enjoying a rich social life.

Despite broad theoretical and biomedical interests in understanding this sociality-fitness nexus, researchers have mainly focused on a single physiological pathway, based on exposure to chronic "social stress" in individuals experiencing poor social environments triggering detrimental neuroendocrine and immune responses (the "social causation" hypothesis; [3,14,15]). Indeed, evidence in laboratory animals indicate that social conditions that promote chronic stress, such as social isolation or the removal of a social companion, predict dysregulation of the hypothalamic-pituitary-adrenal (HPA) axis, changes in signaling in the sympathetic nervous system, elevated cortisol production[14,16,17], as well as immune dysregulation and chronic inflammation[18–20]. Furthermore, when experienced over the long-term, chronic stress seems to predispose individuals to a range of illnesses[21–23] and shorten life span[24,25]. Research on the proximate mechanisms governing phenotypic responses to early life social adversity has likewise been largely limited to studies on glucocorticoids (GCs) [26–28], especially because the HPA axis undergoes important programming in early life and appears particularly sensitive to social perturbations during this specific time window[5,29,30].

Chronic social stress, as a mediating physiological pathway in the sociality-fitness nexus, shows, however, some limits in its explanatory power. From an evolutionary perspective, selection should not maintain a physiological response routinely challenging fitness. Some researchers have argued that animals in their natural environments are unlikely to experience chronic social stress to the degree that it could shorten lifespan[31,32]. In addition, the best evidence for social causation between chronic stress and fitness mostly comes from biomedical research in laboratory animals[3,24]. Comparatively little research has been undertaken in natural populations, and this body of work lead to mixed outcomes with a large majority of non-significant or small effect sizes[31,33]; but see[25]). Exposure to chronic stress is therefore unlikely to explain alone the pervasive sociality-fitness nexus observed in nature.

In parallel, the last decade of research has revealed how microbiomes – the abundant and diverse communities of bacteria, archaea, fungi, and viruses living in and on eukaryotic hosts – participate to the regulation of virtually all aspects of host biology[34]. Symbiotic microbes contribute importantly to host nutrition and metabolism[35,36], educate and modulate the host immune system[37–40], protect against pathogen infections[41–43], and are involved in numerous endocrine and central nervous system signaling pathways[44–46]. In addition, some recent findings have highlighted the extent to which microbial diversity and composition are shaped by the social environment (recently reviewed in [47]), along with phylogeny, diet, habitat, host traits and genetics[48–53]. Here, we refer to the "social microbiome" as the microbial communities that are transmitted via two distinct mechanisms: (1) a vertical transmission of microbial taxa directly from parent to offspring[54] and (2) an horizontal transmission mediated by direct physical contacts or shared environmental surfaces with conspecifics[47] (**Figure 1A**).

Combined together, those findings open a new avenue of research on the microbial pathways through which sociality could contribute to health and fitness outcomes in nature. Although poorly considered in naturalistic studies, previous researchers have argued that the social transmission of beneficial gastro-intestinal (GI) symbionts might mitigate the enhanced risk of pathogen infection in social species, constituting an underappreciated benefit of group living[55–57]. In a recent review, Amato et al.[58] highlighted how the GI microbiome is an important and underappreciated pathway by which differences in social, political, and economic factors could contribute to health inequities in humans. Here, we aim to extend this previous theoretical framework to propose microbiomes as a missing physiological pathway at the very origin of the relationship between sociality and fitness in wild mammals. We propose that the quality of the social environment, both in early life (via mother-to-infant microbial transmission) and adulthood (via socio-sexual relationships with conspecifics), strongly contributes to the diversification of individual microbiomes – at multiple body sites – and to the transmission of beneficial microbial taxa among individuals living in groups. In turn, these more diverse, stable, and valuable microbiomes are predicted to directly improve host development, nutrition, metabolism, immune and neuroendocrine functions, and above all, pathogen resistance throughout lifespan, and might directly explain how social relationships translate into massive positive fitness outcomes in nature. We thus propose an alternative, although non-exclusive physiological mechanism to the "chronic stress" hypothesis deserving further empirical investigations and based on the diversification and social transmission of health-promoting microbes that has great potential to mediate the relationship between each of the three aspects of the social environment that are related to individual fitness.

**Microbiomes impact a wide range of host phenotypes**

In mammals, microorganisms start colonizing the host immediately during and after birth and assemble into ecological communities that quickly change and evolve until reaching a stable, but nonetheless dynamic, microbial composition after weaning[54,59–62]. The trajectory of microbial maturation in early life[63–65] and the microbiome diversity and composition during adulthood[63–65] are both highly variable between individuals. Such inter-individual differences in microbial composition influence a variety of host phenotypes[34,66], including those related to digestion, immunity, behavior, and phenotypic development[46,67,68] and are likely determining host phenotypic plasticity and fitness in wild animals[69].

The GI microbiome, reaching particularly large microbial density in the lower intestinal tract, has been the most studied so far and is involved in many aspects of host biology[68,70,71]. Intestinal microbes metabolize otherwise inaccessible dietary substrates and produce short-chain fatty acids that are used as an additional source of energy for the host[72,73], supply essential vitamins, detoxify plant secondary compounds[74], and participate to important digestive processes such as insulin secretion, maintenance of glucose homeostasis and lipid absorption[75]. They further program several aspects of host energy metabolism, such as the regulation of fat storage[76–78]. In wild mammals, adaptive shifts in GI microbial composition with environmental variation is increasingly recognized as an important physiological mechanism through which animals adapt to seasonal changes in diet and maintain their energy balance during challenging periods[79–85]. This dietary and metabolic flexibility provided by GI microbes influence host phenotypic plasticity and life history traits[34,86,87] and likely participate to host local adaptation and evolution[66,88,89].

In addition, mounting evidence suggests that the GI microbiome strongly affects host resistance and tolerance to numerous pathogenic bacteria, viruses, fungi and parasites[41,42,90–92]. The GI tract is a mucosal surface constantly exposed to the external environment and, as such, has developed elaborated innate and adaptive immune responses to prevent pathogen invasion[37]. GI microbes contribute directly to pathogen resistance by (i) outcompeting pathogens for nutrients and space, (ii) producing antimicrobial and antiviral peptides inhibiting pathogenic growth[41,42,90,91], and (iii) stimulating mucus production and ensuring the integrity of the intestinal barrier[93,94]. Experiments on laboratory animals devoid of GI microbiota ("germ-free") have clearly demonstrated the critical role of the GI microbiome in the resistance to acute bacterial and viral infections[95]. Germ-free or antibiotic-treated mice, for instance, show extremely poor immune resistance and higher mortality when infected by a variety of enteric bacteria (e.g. *Shigella flexneri, Listeria monocytogenes*, *Salmonella enterica*)[96] or virus (e.g. Influenzas A virus)[97]. GI microbes also participate indirectly to pathogen resistance by engaging in a cross-talk with the host immune system upon pathogen detection[43,92]. In particular, the GI microbiome trains the host immune system in early life[46,98,99] and modulates the host immune response later in life, notably by directing the differentiation of the pro- and anti-inflammatory responses following infection[37,38,41,91]. During helminth and protozoa infections, for example, the GI microbiome decreases the pro-inflammatory host immune response, which in turn promotes host tolerance to parasites[100–102]. Commensal microbes thus participate to the equilibrium between inflammation and homeostasis in the gut[38,39].

The functions of other microbial communities have been far less studied but emerging evidence also points to their important role in modulating host immunity[95]. Microbes from the upper respiratory tract shape host susceptibility to several respiratory viral infections[103,104]. For example, *Staphylococcus aureus*, a normal inhabitant of the respiratory mucosa, protects against influenza-mediated lethal inflammation in wild-type mice compared to germ-free mice[105]. The skin microbiome has been associated to infection susceptibility by the fungus *Pseudogymnoascus destructans* causing the white-nose syndrome in North American hibernating bats[106,107]. Resistant individuals harbored a more diverse and abundant host-associated bacterial and fungal community on the skin compared to susceptible bats, and some specific strains were directly found to inhibit the growth of the pathogenic fungus *in vitro*[106,108]. In humans, the female genital microbiome is commonly dominated by the genus *Lactobacillus* that secrete lactic acid and hydrogen peroxide that create an important barrier against sexually-transmitted bacterial (e.g. *Chlamydia trachomatis*, *Gardnerella vaginalis*) and viral (e.g. Human immunodeficiency virus 1, Human herpes simplex virus 2) infections, respectively[109–111]. Some *Lactobacillus* strains seem to be more protective than others and their carriers display a lower susceptibility of acquiring sexually-transmitted diseases[111,112]. Overall, pathogen protection conferred by different microbiomes has been proposed as the main evolutionary advantage for the host to tolerate such dense microbial communities[92].

**Diversified and stable microbiomes are highly beneficial**

In general, more diverse (i.e. higher taxonomic richness) and stable (i.e. the capacity of the community to retain its similarity in composition in response to disturbances) microbial communities are associated with better metabolism functioning[34,113,114], better pathogen resistance[56,115] and optimal health status[113,116,117] for the host. More diverse microbial compositions are thought to use more completely the nutrients and space available at the mucosal surfaces and give less opportunities for pathogens to invade[92,118]. Microbial diversity also increases resilience to community perturbations and promotes microbial stability, by creating functional redundancy between microbes and preserving functional capacity in hosts in case of microbial lineages extinctions[117,119]. Highly diversified GI microbiome, and in particular an increase in the prevalence of rare taxa at the expense of core taxa during old ages, has been related to pattern of healthy aging and longevity in humans[120–122]. By contrast, less diverse and more instable microbial communities generally increase host susceptibility to bacterial and viral infections[38,116,123] and are often associated with poor health outcomes[124]. In humans, depletion or perturbations in microbial communities – due to e.g. sanitation, Western diets and antibiotics or vaccination– have been linked to numerous metabolic and immune disorders including obesity, type 1 diabetes, asthma, atopy, inflammatory bowel disease (IBD) and cardiovascular diseases[116,125–128]. Mechanisms to maintain microbial diversity and stability are thus likely to be under strong directional selection in nature.

**Social contacts promote microbiomes' diversity and stability**

Meanwhile, a growing literature shows that characteristics of the social environment, in particular how individuals are organized in social networks, act as conduits for microbial exchanges within a host population. Vertical transmission from mother to offspring during and just after birth is believed to seed and influence durably both the diversity and composition of the offspring's microbiome[54,129,130], which then continues to be shaped by social contacts with conspecifics later in life through horizontal transmission[47].

*Mother-to-infant vertical transmission.* Mothers are one of the most important reservoirs of microbes for their offspring in early life[54,129–132]. Such vertical transmission is thought to be particularly strong in mammals due to viviparity and prolonged periods of lactation and post-weaning maternal care[48]. During parturition, infants are initially inoculated by their mothers' vaginal, fecal and skin microbiomes[63,130,133,134]. Following birth, microbial transmission from the mother continues with milk microbes' ingestion during nursing[60,135–137] and GI microbes transmission mediated by body contacts[48]. Milk glycans (i.e. oligosaccharides) further stimulate the growth of beneficial microbes and guide the succession of microbial assemblage in the GI tract of offspring[138–140], as well as providing protection against invading pathogens[141,142]. Maternal microbial signature in the infant gut often persist for long period of time because the first taxa to establish benefit from priority effects and are better adapted to the gut ecological niche compared to other environmental microbes[60,130,143]. Thus, mother-to-infant vertical transmission strongly shapes the diversity and composition of the infant gut microbiota, and increasing evidence show that such effect persists post weaning[53,144,145], likely via preferential social bonds between mother and offspring dyads in group-living species[47,48].

*Socio-sexual interactions and microbial exchanges.* During adulthood, microbiomes are also largely transmitted via social and sexual contacts between conspecifics, either by direct physical contacts (e.g. grooming, copulations), shared surfaces, or coprophagy[47,146]. This literature – though highly biased towards the skin and GI microbiomes – indicates that individuals living in the same social group exhibit higher similarity in their microbial communities (e.g. Egyptian fruit bats, *Rousettus aegyptiacus*:[147]; yellow baboons, *Papio* *cynocephalus*:[148,149]; chimpanzees, *Pan troglodytes*:[150]; red-bellied lemur, *Eulemur rubriventer*[151]; Verreaux's sifaka, *Propithecus verreauxi*:[152]; humans:[153,154]). Furthermore, individuals with larger social networks, or those benefiting from a rich social environment, harbor, on average, more diverse microbiomes than those that are poorly socially integrated[47,148,152,155,156]. At the dyadic level, pairs of individuals that engage in more frequent social[47,81,156] and sexual[157–159] interactions also exhibit more similar microbial communities. In yellow baboons, for instance, individuals that spend most of their time grooming each other have the most similar GI microbiomes (about 15-20% of microbial taxa are shared preferentially), after controlling for kinship and shared diet[148]. In kittiwakes (*Rissa tridactyla*), breeding pairs exhibit higher similarity in their cloacal communities[158]. Importantly, the "socially-structured" microbial taxa are preferentially found in some taxonomic groups such as anaerobic GI microbes, transmitted through intimate physical contacts[47]. In baboons, for example, these socially-shared microbial strains include the families *Bifidobacteriaceae*, *Coriobacteriaceae*, *Veillonellaceae* that have all been linked to beneficial health effects in humans[160,161]. The duration and intimacy of sexual and social contacts are therefore expected to influence both the quantity and taxonomic composition of the shared microbial taxa[47].

**The "social microbiomes" likely mediate the effects of early life adversity, social integration and social status on fitness**

Early life microbial colonization and diversification are increasingly recognized as central processes shaping infant's growth and phenotypic development[162–166], the maturation and education of the immune system[40,98,167], as well as neurodevelopment[46,168,169], thus leading to profound health and fitness consequences across lifetime[98,170–172]. As emphasized above, maternal vertical transmission strong shapes the trajectory of microbiome maturation and is likely to play an important role in offspring's initial and future health. Numerous clinical studies in humans, for instance, have shown how perturbations in the normal pattern of mother-offspring vertical transmission during the peri-natal period (e.g. Cesarean section, maternal antibiotic use, or formula feeding) can translate into metabolic- and immune-associated disorders in offspring, that often persist until childhood and adulthood, such as type 1 diabetes, obesity, IBD, or asthma[134,170,171]. In rodents, experimental manipulations have shown that maternal stress during pregnancy causally alters vaginal microbiota composition, and in turn, vertical transmission of this dysbiotic community results in impaired metabolic, immunologic and neurodevelopmental functions in the neonate[173–178].

Social adverse conditions challenging the initial vertical transmission of maternal microbes could therefore strongly mediate the relationships between early-life social adversity and fitness outcomes pervasively observed in nature[28,179] (**Figure 1B**). For example, low-ranking or poorly socially integrated mothers may carry and transmit less diversified milk or GI microbiomes to their infants, potentially explaining why their offspring usually grow slower[180–182] and display shorter lifespan[4–6,183–185] than those born to high-ranking mothers or to those benefiting from a rich social life. Recent studies in mammals have shown for example that maternal traits, such as parity (i.e., the number of times a female reproduced), are associated with difference in offspring's GI composition[163,186,187] and translate into variation in the speed of offspring's GI microbiome maturation[163,186]. Similar mechanisms could mediate the relationship between maternal social environment and offspring fitness. In addition, early maternal death – even post-weaning – may preclude an appropriate long-lasting maternal source of microbes, again leading to little diversified and beneficial microbial taxa in offspring (**Figure 1B**), possibly explaining the major detrimental effects of early maternal loss on offspring development and fitness.

Highly socially-integrated or high-ranking individuals are generally exposed to numerous socio-sexual partners and have more choice over their partners[188–190]. These highly-integrated or/and high-ranking individuals are thus likely to harbor more diverse and stable microbiomes throughout their life because of their numerous social partners[47], but could also benefit from better microbial strains if they select socio-sexual partners depending on their microbial communities. Both diversified and beneficial microbial communities would directly improve nutrition, physiology and, above all, pathogen resistance in these social individuals, and ultimately explain their improved health, longevity and reproductive success[56,58,191] (**Figure 1B**). Yet, evidence of microbially-mediated health and protective benefits acquired from the social environment remains elusive so far, and is currently restricted to eusocial insects. In cockroaches and termites (*Dictyoptera* sp., *Cryptocercus punctulatus*), coprophagy is essential to newly hatched nymphs to acquire cellulolytic intestinal symbionts from conspecifics and digest their diet[192,193]. In honey bees (*Apis mellifera*), GI microbes acquired from nestmates in early life reduce individual susceptibility to the protozoa *Lotmaria passim*[194]. Similarly, socially-transmitted GI microbial taxa protects bumble bees (*Bombus terrestris*) from infection by the trypanosomatid *Crithidia bombi*[195]. In group-living mammals, social behavior may have evolved to buffer against the transmission of pathogenic microbes (e.g. via avoidance of sick individuals or careful socio-sexual partner choice[196–198]), while facilitating sharing the beneficial ones. Such microbially-mediated effects of the social environment on fitness is expected to be strong especially (i) when beneficial symbionts cannot be obtained from the environment, such as anaerobic GI microbes that can only be transmitted through physical contacts[47], and (ii) when repeated inoculations of endosymbionts are necessary to establish or maintain a functioning assemblage[55]. Empirical investigation needs thus to be undertaken in social mammals to elucidate these pathways.

**Effects of the GI microbiome on physiological stress**

The GI microbiome engages in a bidirectional communication with the brain through neural, endocrine and immune pathways (the "gut–brain axis")[45,199,200]. Intestinal microbes produce a variety of neurotransmitters (e.g. gamma-aminobutyric acid GABA, serotonin) and metabolites (e.g. short chain fatty acids) that interact with the HPA axis, influence GCs synthesis, and participate to signaling pathways in the brain[45,200–203]. In particular, the GI microbiome is tightly linked to the activity of the HPA axis and emerges as a key player in the physiology of the stress response [202,204–206]. Laboratory studies have demonstrated the extent to which the development and regulation of the HPA axis is controlled by the microbiome and, in turn, how the GI microbiome responds to signals sent by the HPA axis[202,203,205,207]. In adult mice, for instance, *Enterococcus faecalis* promotes affiliative social behavior during encounters with a novel individual by dampening the HPA-axis-mediated production of corticosterone and by suppressing overactive stress response[208]. In early life, the programming of the HPA axis by the GI microbiome shapes the overall stress reactivity over lifespan: germ-free mice generally display HPA hyper-activity and produce abnormal anxiety-like behaviors[202,205,207,209–211]. Maternal separation or antibiotic exposure that translate into alterations of early microbial compositions can result in the long-term modulation of stress-related physiology and behavior[212]. By contrast, microbial recolonization can normalize adrenal responses but only within a critical window during early development[210,213].

Our understanding of these processes remains nonetheless limited in natural populations. The few empirical studies so far show that elevated GC levels, resulting from environmental and/or social stressors, can actively disrupt the diversity or composition of the GI microbiome[214–219] which is, in turn, generally associated with negative health outcomes for the host[45,202,204,205]. Since the GI microbiome and the HPA axis are engaged into a bidirectional communication, the causality can be often difficult to establish. These studies nonetheless reveal that the microbiome is an integrant part of the stress response that need to be considered when investing the sociality-fitness nexus.

**Conclusions and future directions**

The social transmission of microbes has the potential to confer fitness benefits in ways that could explain the sociality-fitness nexus, to a similar – or even higher – extent than chronic social stress. As discussed above, microbiomes regulate virtually all aspects of host physiology, with pervasive effects on endocrine, metabolism and immune functions of organisms during development and adulthood[220]. More than 90% of the current studies have focused on the GI microbiome, but microbes at multiple body sites (buccal, nasal, skin, respiratory, or genital microbiomes) appear as promising candidates to understand inter-individual differences in reproductive success and lifespan in nature and deserved further investigation. Additionally, microbially-mediated pathogen protection has the potential to explain large effect size of the social environment on fitness because pathogens exert one of the most important selective pressures on their hosts in nature. To interpret the widespread and pervasive sociality-fitness nexus reported in wild social mammals, we propose here that a far-reaching part of the explanatory power traditionally attributed to chronic social stress might in fact result both from the diversification of these communities via socio-sexual contacts, and from the social transmission of beneficial microbial communities.

Long-term studies in wild social mammals are particularly well suited to investigate this alternative mechanism. First, the social environments of nonhuman mammals are simpler than the ones experienced in humans and measures of early life adversity, social integration and social status have been well characterized and are strongly linked to individual fitness[3]. Furthermore, most studies of host-microbiome relationships have been conducted either in (western) humans, where experimentations are limited for ethical reasons, or in laboratory (germ-free) rodents. The continued utilization of these models restricts our understanding of the microbiome as a complex ecosystem governed by eco-evolutionary processes. Wild mammal microbiomes are in less altered states than those found in contemporary humans that experience modified selective pressures due to modern medicine, industrial diet and lifestyle[86]. There is an urgent need to incorporate studies of non-model species living in ecologically realistic environments. Species living in large groups, with hierarchical relationships and well differentiated and stable social and sexual interactions, are particularly interesting because such setting may generate important inter-individual differences in microbial social transmission. Species characterized by long developmental periods with associated maternal effects are also particularly interesting to investigate in such contexts. Finally, focusing on animals experiencing high parasitic pressures, with consequences on individuals' behaviors and fitness, is of outmost importance to investigate the contribution of microbially-meditated pathogen protection as a mediating mechanism explaining the relationship between sociality and fitness.

## ACKNOWLEDGMENT

We thank Dr Lauren Petrullo for providing feedbacks on an earlier version of this manuscript.

**Figure 1.** Microbially-mediated effects of social environment on fitness in nature.

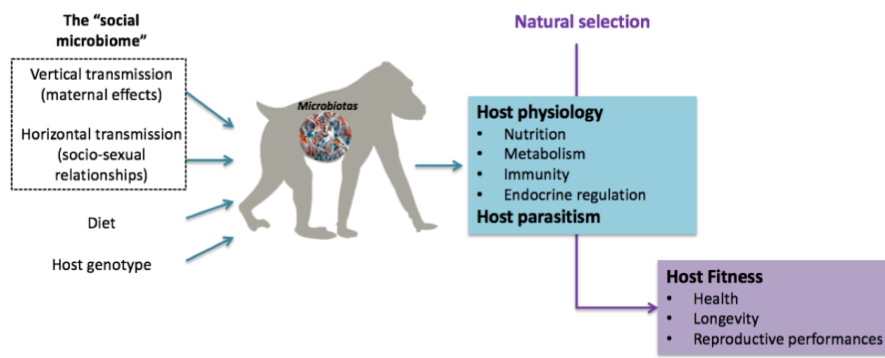

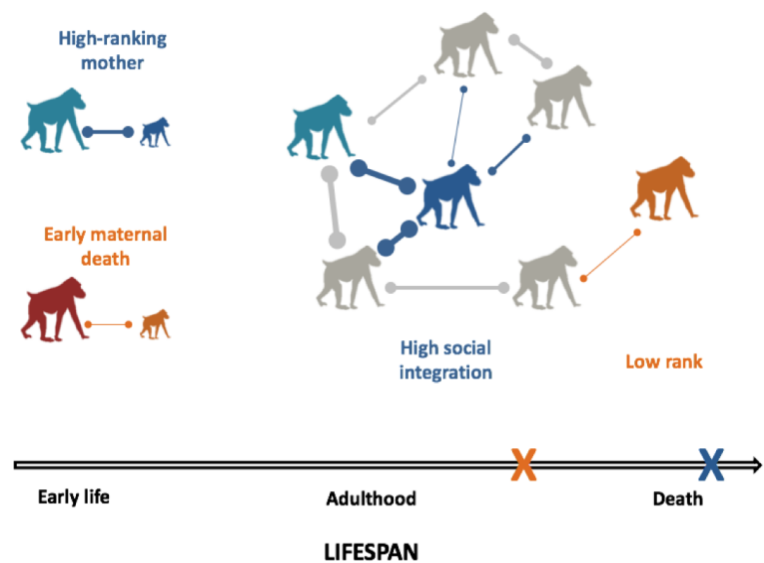

(A) Microbiotas are dynamic communities shaped by multiple factors throughout an individual's life. Vertical and horizontal transmissions strongly participate to the diversification of individuals' microbiotas, which in turn promote host physiology and pathogen resistance, and is likely to translate into positive fitness outcomes. Adapted from [86]

(B) Theoretical developmental trajectories of two individuals, from birth to death. The dark blue individual benefits from two positive social effects (high-ranking mother and high social integration) while the orange individual experiences two adverse social conditions (early maternal death and low social rank), resulting in late vs. early death respectively (colored crosses). The light blue mother survives until her offspring's adulthood and constitutes an important social partner. The thickness of the links represents the strength of the social bonds that act as conduits for microbial exchanges.